\documentclass[12pt]{article}

\usepackage{amssymb,amsmath,amsfonts,eurosym,geometry,ulem,graphicx,caption,color,setspace,sectsty,comment,footmisc,caption,pdflscape,subcaption,dsfont,natbib,hyperref
}
\usepackage{multirow}
\usepackage{array,sgame,istgame }
\usepackage{fouriernc}
\usepackage[flushleft]{threeparttable}
\usepackage{nicematrix}

\usepackage{algorithm,algpseudocodex} 
\usepackage{adjustbox}
\usepackage{bm}

\newcolumntype{L}[1]{>{\raggedright\let\newline\\arraybackslash\hspace{0pt}}m{#1}}
\newcolumntype{C}[1]{>{\centering\let\newline\\arraybackslash\hspace{0pt}}m{#1}}
\newcolumntype{R}[1]{>{\raggedleft\let\newline\\arraybackslash\hspace{0pt}}m{#1}}

\begin{document}

\begin{titlepage}
\title{
\fontsize{18pt}{18pt}\selectfont
Strategic Information Disclosure in Algorithmic Pricing
}
\author{Chengcheng Wang \and 
Zexin Ye\thanks{Chengcheng: cwang111@arizona.edu, Department of Economics, The University of Arizona. Zexin: ye.754@osu.edu, Department of Economics, The Ohio State University. We are grateful to Yaron Azrieli, Andreas Blume, Inga Deimen, Paul J. Healy, Aldo Lucia, Jim Peck, Matthijs Wildenbeest, Huanxing Yang, and Lixin Ye for their insightful comments.  
We appreciate the valuable feedback from the OSU Theory/Experimental Reading Group and the University of Arizona Theory Lunch.  
All errors remain our own.}}
\date{\today}
\maketitle

\begin{abstract}
\noindent
As firms increasingly adopt AI-powered pricing algorithms, a key and urgent policy concern is how to regulate the potential algorithmic collusion. This paper approaches the regulatory question through the lens of information design and examines how different disclosure rules, committed to by a third-party intermediary, shape learning outcomes when firms delegate pricing to Q-learning algorithms under stochastic demand.
We analyze three disclosure rules: no disclosure, full disclosure, and upper censorship. Upper censorship, which truthfully reveals low-demand states while pooling high-demand ones, delivers higher profits than full disclosure, consistent with theoretical predictions. However, we uncover a profit reversal: when the discount factor is high, no disclosure yields higher profits than full disclosure, whereas when the discount factor is low, full disclosure performs better. This pattern is exactly the opposite of what classical collusion theory predicts.
Overall, these findings show that Q-learning agents respond systematically to the information structure and further suggest that restricting information sharing may backfire when algorithms are sufficiently patient, highlighting the need to reassess regulatory approaches in AI-mediated markets.
\\
\vspace{0in}
\vspace{0in}


\bigskip
\end{abstract}
\setcounter{page}{0}
\thispagestyle{empty}
\end{titlepage}
\pagebreak \newpage

\doublespacing

\section{Introduction}
Many recent studies show that AI-powered pricing algorithms can autonomously learn to sustain supracompetitive profits in repeated interactions, even without explicit communication or human intervention (for a review, see \citet{assad2024algorithmic}). This emerging form of algorithmic collusion poses a fundamental challenge for antitrust enforcement, which traditionally relies on detecting a meeting of the minds or other forms of explicit communication between firms \citep{h2018ac,harrington2025critique}. Thus, there is an urgent need to understand how algorithmic pricing should be regulated.

In this paper, we study information design as a tool for regulating algorithmic pricing. Specifically, we examine how different information disclosure rules shape the learning outcomes of Q-learning algorithms. We focus on environments with third-party information disclosure, which has become a central issue in both academic research and policy debates\citep{harrington2025economic}.\footnote{Individual firms typically rely on third-party intermediaries to obtain information about market conditions, while these intermediaries possess a comparative advantage in aggregating and estimating such information \citep{harrington2025critique}.}
For example, RealPage collects extensive rental data and provides pricing recommendations to landlords. Empirical evidence shows that the use of such information can raise rental rates and tenant expenditures\citep{calder2024algorithmic}.

To analyze how such information structures shape algorithmic pricing, we consider a market environment with uncertainty generated by stochastic demand shocks. Firms do not observe the realized demand state and therefore cannot condition their prices on it. In contrast, a third-party intermediary observes the true demand and commits to an information disclosure rule that determines what signal is released to firms. Within this framework, we examine three disclosure rules: no disclosure, full disclosure, and a selective disclosure known as upper censorship, which has been shown to be optimal for maximizing collusive profits under affine demand functions \citep{sugaya2025collusion}. Firms delegate pricing decisions to Q-learning algorithms, which incorporate the demand signals released by the third party.

Our results reveal three main findings.
First, Q-learning agents respond systematically to the information structure of demand signals: different disclosure rules generate distinct profit paths across the discount factor $\delta$. 
Second, under upper censorship, varying the truth-telling probability $\rho$ generates sharply different profit trajectories, thereby expanding the set of attainable profits within this disclosure rule. Consistent with theoretical predictions, upper censorship also strictly dominates full disclosure: for every $\delta$, the set of attainable profits under upper censorship contains that under full disclosure.
Third, and most strikingly, we uncover a robust profit reversal between no disclosure and full disclosure. When $\delta$ is low, full disclosure yields higher profits, whereas when $\delta$ is high, no disclosure produces the higher profits—exactly the opposite of the classical theoretical prediction. This reversal persists across demand environments (both linear and logit), indicating a fundamental divergence between Q-learning dynamics and equilibrium-based comparative statics.

This paper also delivers important policy implications. It provides the first systematic analysis of how information disclosure affects algorithmic pricing, showing that Q-learning algorithms respond systematically to the structure of information disclosure. This implies that information design can serve as a regulatory instrument in algorithmic pricing, allowing regulators to choose different disclosure rules to achieve different policy objectives.
In addition, our finding of profit reversal directly contradicts the comparative statics implied by theory. This result highlights a crucial policy message: restricting information sharing may backfire in AI-mediated markets, potentially strengthening rather than weakening collusion, consistent with the concerns raised by \citet{harrington2025critique}.

The rest of the paper is organized as follows. Section \ref{sec:literature} reviews the related literature. Section \ref{sec:theory} introduces the economic model, the information disclosure rules, and the optimal collusive outcomes under each disclosure rule. Section \ref{sec:algorithm} describes the Q-learning algorithms and the state representation under each disclosure rule. Section \ref{sec:results} reports the main simulation results. Section \ref{sec:robust} presents a series of robustness checks. Section \ref{sec:conclusion} concludes.

\section{Literature Review}\label{sec:literature}
This paper relates to two strands of literature.
First, a large theoretical literature studies the sustainability of collusion under different market structures and information environments. Classic work such as \citet{gp1984} and \citet{rs1986} analyzes repeated price competition with stochastic demand. Subsequent research incorporates informational frictions, with \citet{mt2019demand} and \citet{ow2021demand} examining how public demand signals of varying precision affect firms’ ability to sustain collusive outcomes. In markets where information is mediated by a third party, \citet{harrington2025hub} analyzes hub-and-spoke collusion, in which a central information aggregator facilitates coordinated behavior among competing firms.
\citet{kamenica2011bayesian} initiate the information design literature by showing that strategic information disclosure can systematically alter equilibrium outcomes. Within a collusion context, \citet{sugaya2025collusion} show that the optimal information disclosure rule is upper censorship under affine demand. In our paper, we connect information design with algorithmic pricing, evaluate how Q-learning algorithms respond to them.

Second, a rapidly growing literature shows that pricing algorithms can autonomously learn to sustain supracompetitive outcomes. \citet{calvano2020} show that Q-learning agents in a repeated Bertrand game with logit demand learn to set supracompetitive prices supported by reward-punishment strategies. \citet{klein2021} finds that Q-learning algorithms collude in dynamic sequential pricing, and \citet{calvano2021} demonstrate that tacit collusion can arise even under imperfect monitoring with unobserved demand shocks. More recent research explores alternative learning rules, memory structures, and market environments \citep{banchio2022artificial,fgs2024,ye2025algorithmic}, as well as applications to platforms and auctions \citep{jrw2023,bs2022}. Related work also examines the policy implications of algorithmic collusion, highlighting potential challenges for traditional antitrust enforcement \citep{harrington2025critique}. This paper contributes to this literature by embedding Q-learning agents in an environment with third-party information disclosure and by systematically comparing how different disclosure rules shape profit levels.


\section{Theoretical Part}\label{sec:theory}
\subsection{Model}
Two firms participate in an infinitely repeated Bertrand competition game, each producing homogeneous goods that are perfectly substitutable. In each period, an i.i.d. demand shock occurs, taking value $\theta_t \in \{\theta_L,\theta_H\}$ with equal probability, where $L$ corresponds to the low demand state and $H$ corresponds to the high demand state, i.e., $\theta_L < \theta_H$. The firms cannot directly observe the realization of the demand shock. Instead, a third party observes the demand state in each period. The third party commits to an information disclosure rule and releases a demand signal $m_t \in \{l,h\}$ in each period according to this rule. Upon observing the signal, the agents simultaneously set prices. The agent charging the lower price captures the entire market demand, while the market is split evenly in the case of a tie.
The demand function for firm $i$ in period $t$ is
$$
D_{it}(p_{i t}, p_{-i t};\theta_t)= 
\begin{cases}

\theta_t-p_{i t} & \text { if } p_{i t}<p_{-it} \\ 

\dfrac{ \theta_t-p_{i t}}{2} & \text { if } p_{i t}=p_{-it}\\ 

0 & \text { if } p_{i t}>p_{-it}

\end{cases}
$$
\noindent Correspondingly, the realized period payoff for firm $i$ given $\theta_t$, is $\pi_{it}=p_{it}  D_{it}(p_{i t}, p_{-i t};\theta_t)$, where marginal cost is normalized to zero. The third party's interest is assumed to be aligned with firms.

Note that the demand signal $m_t$ does not affect the demand function directly; instead, it shapes firms’ beliefs about the true demand state, and therefore influences their pricing strategies. Instead, upon observing the signal $m_t$, firms form a belief
$
\mu_t=\operatorname{Pr}\left(\theta_t=\theta_H \mid m_t\right)
$
and base their pricing decisions on the corresponding expected profit
$$
\mathbb{E}\left[\pi_{i t}\left(p_{i t}, p_{-i t} ; \theta_t\right) \mid m_t\right]=\sum_{\theta_t } \operatorname{Pr}\left(\theta_t \mid m_t\right) \pi_{i t}\left(p_{i t}, p_{-i t} ; \theta_t\right)
$$
Each firm chooses prices to maximize the discounted expected profits 
$$
\sum_{t=0}^{\infty} \delta^t \mathbb{E}\left[\pi_{i t}\left(p_{i t}, p_{-i t} ; \theta_t\right) \mid m_t\right]
$$
where $\delta$ is the common discount factor.

We focus on symmetric subgame perfect equilibria supported by grim-trigger strategies. In such equilibria, the collusive outcome must satisfy the standard incentive compatibility constraint: the deviation gain must be no larger than the discounted continuation loss induced by punishment.

\subsection{Information Disclosure Rules}
We now introduce the information disclosure rules and, for each rule, provide the intuition for how firms would set prices to sustain the highest possible collusive profit. Technical details are provided in the Appendix \ref{app:techniques}.

\subsubsection{No Disclosure: Unobserved Demand Shocks}
No disclosure arises when the third party either does not observe the demand shock or chooses not to reveal it. As a result, firms cannot distinguish the realized demand state and therefore cannot condition prices on demand. To sustain the highest collusive profit, they must treat every period as if demand were at its expected level and adopt a demand-independent collusive price, specifically the monopoly price evaluated at expected demand.

\subsubsection{Full Disclosure: Observed Demand Shocks}
When the third party fully discloses the demand state, firms observe the realized demand shock. Because deviation incentives are stronger in booms than in downturns \citep{rs1986}, the incentive compatibility constraint binds only under $H$. As a result, sustaining monopoly pricing in both states is feasible only when
$
\delta \ge \delta^c = \frac{2 \theta_H^2}{3 \theta_H^2 + \theta_L^2}
$.
When $\delta < \delta^c$, the monopoly price under $H$ violates the incentive constraint, so firms must lower the collusive price in that demand state, and the optimal price under $H$ is determined by the binding constraint. In contrast, the monopoly price under $L$ is sustainable whenever $\delta > 0.5$, since deviation incentives are weaker in downturns.

\subsubsection{Selective Disclosure: Upper Censorship}
Since full disclosure provides stronger deviation incentives in booms and thus leads to inferior collusive performance (no better than under unobserved demand shocks for a wide range of $\delta$), a natural way to improve collusive performance is to reduce the deviation incentive that arise in booms. One solution, proposed by \cite{sugaya2025collusion}, is \textit{upper censorship}. For intuition, we first explain upper censorship in a continuous-demand setting and then specialize it to the binary case. The idea is to pool all demand states above a cutoff $\theta^{c}$ while truthfully revealing all demand states below it. Upon receiving the pooled demand signal, the conditional expected demand state is $\mathbb{E}[\theta \mid \theta \ge \theta^{c}]$.

\cite{sugaya2025collusion} show that under linear demand, upper censorship is the optimal information disclosure policy, yielding the highest collusive profit among all disclosure rules. The optimal cutoff $\theta^{c}$ is the highest threshold at which the monopoly price evaluated at $\mathbb{E}[\theta \mid \theta \ge \theta^{c}]$ remains incentive compatible. Raising the cutoff increases the pooled expectation and thus the monopoly price, but it also strengthens the deviation incentive, eventually violating the incentive constraint. Lowering the cutoff has the opposite effect: it weakens the deviation incentive and preserves incentive compatibility, but at the cost of pooling lower demand states and reducing collusive profits. The optimal cutoff balances these two forces.

In the baseline, the demand state is binary. Thus, upper censorship takes the following form. When the third party observes $\theta_L$, it truthfully reports $\theta_L$ with probability $\rho$ and instead reports $\theta_H$ with probability $1-\rho$. When it observes $\theta_H$, it always reports truthfully. The optimization problem no longer involves choosing an optimal cutoff $\theta^{c}$ as in the continuous case, but instead choosing the optimal truth-telling probability $\rho^{*}$. The signal structure and posterior are summarized in Figure~\ref{fig:uppercensorship}.

\begin{figure}[h!]
\centering
\begin{subfigure}{0.35\textwidth}
$$
\begin{array}{c|cc}
 & L & H \\ \hline
m=l & \rho & 0 \\
m=h & 1-\rho & 1
\end{array}
$$
\caption{Signal Structure}
\end{subfigure}
\hspace{0.03\textwidth}
\begin{subfigure}{0.35\textwidth}
$$
\begin{array}{c|cc}
 & L & H \\ \hline
m=l & 1 & 0 \\
m=h & \tfrac{1-\rho}{2-\rho} & \tfrac{1}{2-\rho}
\end{array}
$$
\caption{Posterior}
\end{subfigure}
\caption{Upper Censorship in the Binary Case}
\label{fig:uppercensorship}
\end{figure}

\subsection{Theoretical Predictions}
\begin{figure}[h!]
    \centering
    \includegraphics[width=0.99\textwidth]{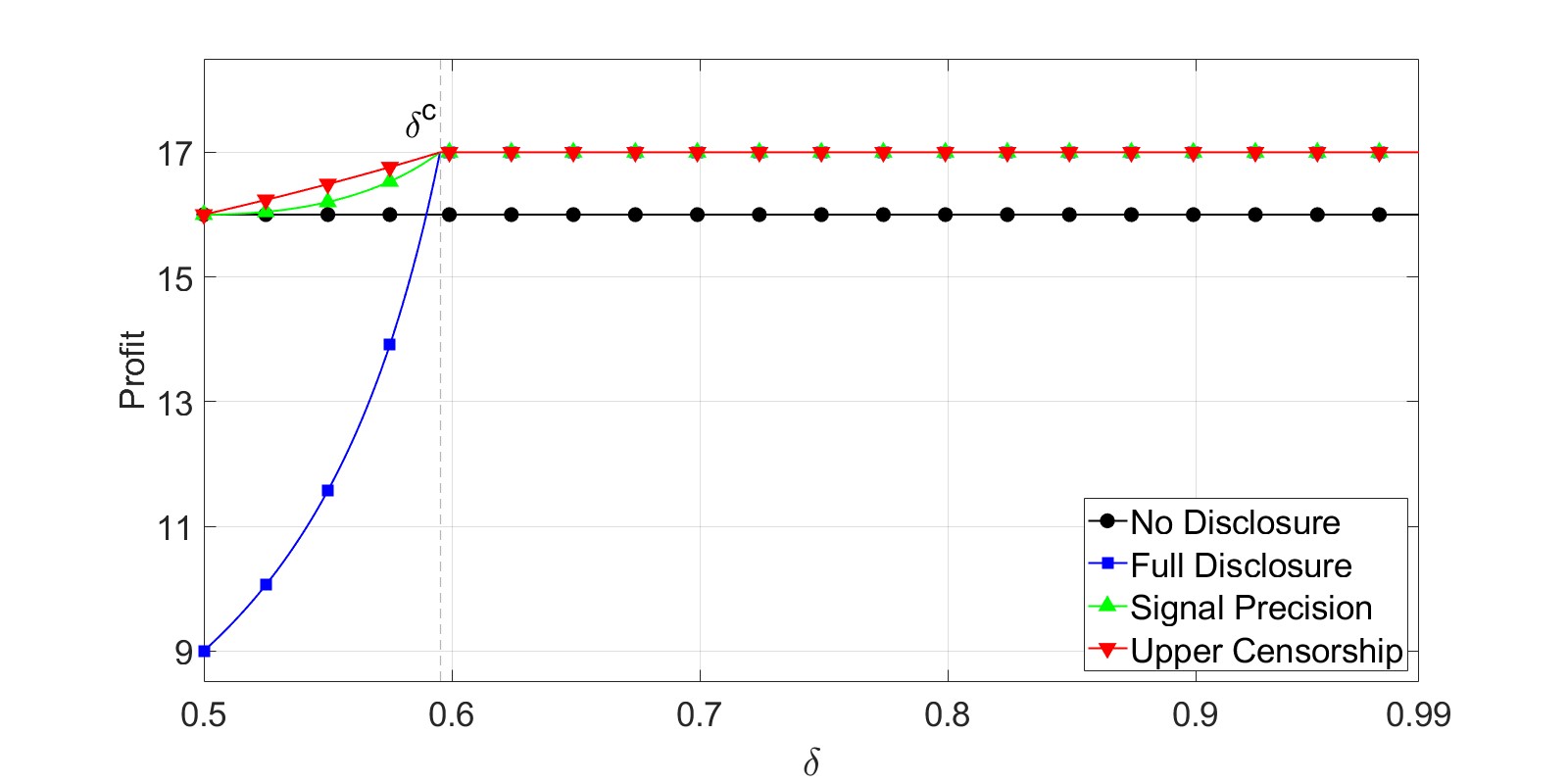}
    \caption{Theoretical Optimal Joint Profits under Different Disclosure Rules}
    \label{fig:theory_profits}
\end{figure}

Figure~\ref{fig:theory_profits} illustrates the predicted optimal (expected) joint profits across $\delta$ under different information disclosure rules, given $\theta_H = 10$ and $\theta_L = 6$.\footnote{For details on the corresponding optimal prices, see Figures~\ref{fig:theory_prices} in the Appendix.} We also report the theoretical benchmark for a second selective disclosure rule, \textit{signal precision}, which will be examined later in robustness checks.

The figure illustrates several theoretical patterns.
Under no disclosure, joint profits remain constant across all $\delta \ge 0.5$ because firms cannot condition prices on demand states, and the optimal collusive price is therefore demand independent.
Under full disclosure, a cutoff discount factor arises. When $\delta$ is above this cutoff, the monopoly prices in both demand states satisfy the incentive constraint, and fully collusive profits are sustainable. When $\delta$ falls below the cutoff $\delta^c$, the collusive price under $H$ violates the incentive constraint, reflecting the stronger deviation incentive in booms documented by \cite{rs1986}. The optimal collusive price under $H$ must therefore be reduced so that the incentive constraint under $H$ binds. For sufficiently low $\delta$, the resulting profit even falls below the collusive profit under no disclosure.
In contrast, upper censorship mitigates the deviation incentive in booms by pooling high demand states. This allows firms to sustain higher collusive prices than under full disclosure when $\delta \in [0.5,\delta^c]$. As a result, upper censorship dominates both full disclosure and no disclosure.

\begin{flushleft}
\textbf{Prediction 1:} 
Upper censorship is optimal in maximizing collusive profits among all disclosure rules, delivering strictly higher profits for all $\delta \in (0.5,\delta^c)$.
\end{flushleft}

\begin{flushleft}
\textbf{Prediction 2:} 
No disclosure delivers strictly higher joint profits than full disclosure for all $\delta \in (0.5,\delta^*)$.
\end{flushleft}

\begin{flushleft}
\textbf{Prediction 3:} 
Full disclosure delivers strictly higher joint profits than no disclosure for all $\delta > \delta^*$.
\end{flushleft}

\section{Q-learning Algorithms}\label{sec:algorithm}
The pricing algorithm we adopt is the Q-learning algorithm introduced by \citet{watkins1989}. This algorithm is widely used in the algorithmic-pricing literature \citep{calvano2020,klein2021,jrw2023} because it is a model-free reinforcement learning method that does not require knowledge of transition probabilities across states. Instead, Q-learning approximates the optimal value and policy functions through pure trial-and-error updates generated by repeated interaction with the environment.

The implementation of Q-learning requires a finite action space. We therefore discretize the continuous price range between the lowest feasible price (the Bertrand equilibrium price at $\mathrm{L}$) and the highest feasible price (the monopoly price at $\mathrm{H}$) into $m$ equally spaced grid points.
The state $s_t$ contains two essential components: the past history and the current demand signal. To maintain computational feasibility, we impose a bounded memory of $K=1$, meaning that Q-learning agents only recall the previous period’s history. Formally, the state is defined as
$s_t = (m_{t-1}, p_{1t-1}, p_{2t-1}, m_t)$.

\subsection{Learning Equation}
The Q-learning algorithm approximates the Q-matrix in an iterative manner. At period $t$, the third party observes the realized demand state $\theta_t$ and then discloses a demand signal $m_t$ according to a committed disclosure rule. Both Q-learning agents receive $m_t$ and form the state $s_t = (m_{t-1}, p_{1t-1}, p_{2t-1}, m_t)$.
Each agent then simultaneously selects its price $p_{it}$ according to the action-selection rule described later. The period payoff $\pi_{it}(p_{1t},p_{2t};\theta_t)$ is realized. At next period $t+1$, after observing the next-period demand signal $m_{t+1}$ derived from $\theta_{t+1}$, agent $i$ updates the corresponding entry $(s_t,p_{it})$ in its Q-matrix using the learning equation:
\begin{equation}
Q_{it+1}(s,p)=(1-\alpha)Q_{it}(s,p)+
\alpha\left[\pi_{it}+
\delta \max _{p^{\prime} \in A} Q_{it}(s^{\prime}, p^{\prime}) 
\right] \label{eq:learningfunc}
\end{equation}
where $ s^{\prime}=s_{t+1} = ( \theta_{t},p_{1t}, p_{2t}, \theta_{t+1} )$. 
The new value $Q_{i,t+1}(s,p)$ is obtained as a convex combination of the previous Q-value and the newly learned experience, which incorporates the current reward and the realized discounted continuation value. All other entries with $s \neq s_t$ or $p \neq p_{it}$ remain unchanged; that is, $Q_{i,t+1}(s,p)=Q_{it}(s,p)$. Thus, only a single Q-value is updated in each iteration, a procedure known as \textit{asynchronous updating}.

The learning rate $\alpha$ controls the weight placed on new information relative to the previous Q-value in the convex update. The discount factor $\delta$ determines the relative importance of the immediate reward compared with the future continuation value.

\subsection{Action Selection}
Each agent’s pricing decision is governed independently by the classic $\varepsilon$-greedy rule:
\begin{equation}
p_{it}\begin{cases}
=\underset{p \in A}{\mathrm{argmax}} \ Q_{it}(s_t,p)  &\text { with the prob. } 1 - \varepsilon_t   \\ 
\sim \text{Uniform}(A)  &\text { with the prob. } \varepsilon_t
\end{cases} \label{actionmode}
\end{equation}
In each period, the agent either selects the price that currently attains the highest Q-value (exploitation, with probability $1-\varepsilon_t$), or draws a price uniformly from the action set (exploration, with probability $\varepsilon_t$).
The exploration rate $\varepsilon_t = e^{-\beta t}$ decreases over time, meaning that the intensity of exploration gradually diminishes as agents accumulate more experience.

\subsection{Simulation Settings}
In the baseline simulation, we set the learning rate to $\alpha = 0.15$ and the exploration parameter to $\beta = 4 \times 10^{-6}$. The action space consists of $m = 11$ equally spaced price points, producing the discrete action set $A = \{0, 0.5, \dots, 5\}$.
The demand state takes two possible values, $\theta_t \in {6,10}$, corresponding to a low demand state ($L$) and a high demand state ($H$), respectively. Under $H$, the competitive price is $p^c_H = 0$ and the monopoly price is $p^m_H = 5$. Under $L$, the competitive price remains $p^c_L = 0$, while the monopoly price decreases to $p^m_L = 3$.

\paragraph{Treatments}
The exact form of the state depends on the information disclosure rule.
Under no disclosure, no signals are released. Thus,
$$
s_t=(\varnothing, p_{1t-1}, p_{2t-1}, \varnothing)
$$
Under full disclosure, the true demand state is fully revealed each period, so the state becomes
$$
s_t=(\theta_{t-1}, p_{1t-1}, p_{2t-1}, \theta_t)
$$
Under upper censorship, the third party discloses a censored demand signal $m_t(\rho)$, where $\rho$ is the truth-telling probability of observing $L$. Accordingly, the state is
$$
s_t=(m_{t-1}(\rho), p_{1t-1}, p_{2t-1}, m_t(\rho))
$$
Because pricing behavior in Q-learning is determined endogenously through learning rather than by solving incentive constraints, the optimal $\rho^*(\delta)$ cannot be derived analytically. We therefore evaluate a discrete grid of values $\rho \in \{0.1,0.2,\dots,0.9\}$.

The discount factor varies over $\delta \in [0.50, 0.99]$ in increments of $0.01$. For no disclosure and full disclosure, each configuration $(\alpha, \beta, \delta)$ is simulated with $1,000$ independent sessions. Under upper censorship, the disclosure parameter $\rho$ introduces an additional dimension, so we run 1,000 sessions for each $(\alpha, \beta, \delta, \rho)$ configuration.

\paragraph{Convergence}
Q-learning is known to converge in single-agent environments \citep{wd1992Q}, but no general convergence result exists for strategically interdependent environments. We therefore adopt the ex-post convergence criterion proposed by \citet{calvano2020}. A simulation is deemed converged when either (i) one billion iterations have been completed or (ii) The optimal strategies for both agents remains unchanged for $100,000$ consecutive periods.\footnote{The optimal strategy for agent $i$ is defined as 
$p_{it}(s)=\underset{p\in A}{\mathrm{argmax}}\, Q_i(s,p)$ for all $s$.}
The simulation stops as soon as either condition is satisfied. 


\section{Results}\label{sec:results}

\subsection{Price Cycle}
After learning is complete, we can derive the limit strategies, defined as the optimal policies implied by the converged Q-matrices:
\begin{equation}
p_i^{*}(s) = \underset{p \in A}{\mathrm{argmax}}, Q_i(s,p), \quad \forall s \in S
\label{optimal_policy}
\end{equation}
Based on the limit strategies, we can represent the induced price dynamics using a directed graph. Each node consists of a demand signal and a price pair, and each directed edge represents a transition from one node to another as prescribed by the limit strategies.\footnote{For example, $v_i = (\theta^i, p_1^i, p_2^i)$ points to $v_j = (\theta^j, p_1^j, p_2^j)$ if, given the state $(\theta^i, p_1^i, p_2^i, \theta^j)$, the optimal prices chosen by both agents are $p_1^j$ and $p_2^j$, respectively.}
Then we use the directed graph to identify the price cycle, which is a key step in analyzing the simulation results. The price cycle is defined as a subgraph that is absorbing and in which every node is visited with strictly positive probability. To obtain the price cycle, we follow the method proposed by \citet{ye2025algorithmic}.
The price cycle has the appealing property that the stationary distribution over its nodes is unique and assigns positive probability everywhere.\footnote{In graph-theoretic terms, the price cycle corresponds to an absorbing strongly connected component (SCC).} Given this stationary distribution, we compute the average long-run prices and profits implied by the price cycle.


Once we compute the average long-run profit, we construct a normalized profit index to measure the degree of collusive outcomes.
We first compute the average joint profit across simulation sessions:
$$
\overline{\Pi}
= \frac{1}{1000} \sum_{k=1}^{1000} 
\left( \pi_1^{(k)} + \pi_2^{(k)} \right)
$$
where $\pi_i^{(k)}$ denotes the expected profit of agent $i$ in session $k$. We then normalize the average joint profit relative to the competitive and monopoly benchmarks:
$$
\Delta
= \frac{\overline{\Pi} - \Pi^{C}}
       {\Pi^{M} - \Pi^{C}}
$$

\subsection{Profit Comparisons}
\begin{figure}[h!]
    \centering
    \includegraphics[width=0.95\textwidth]{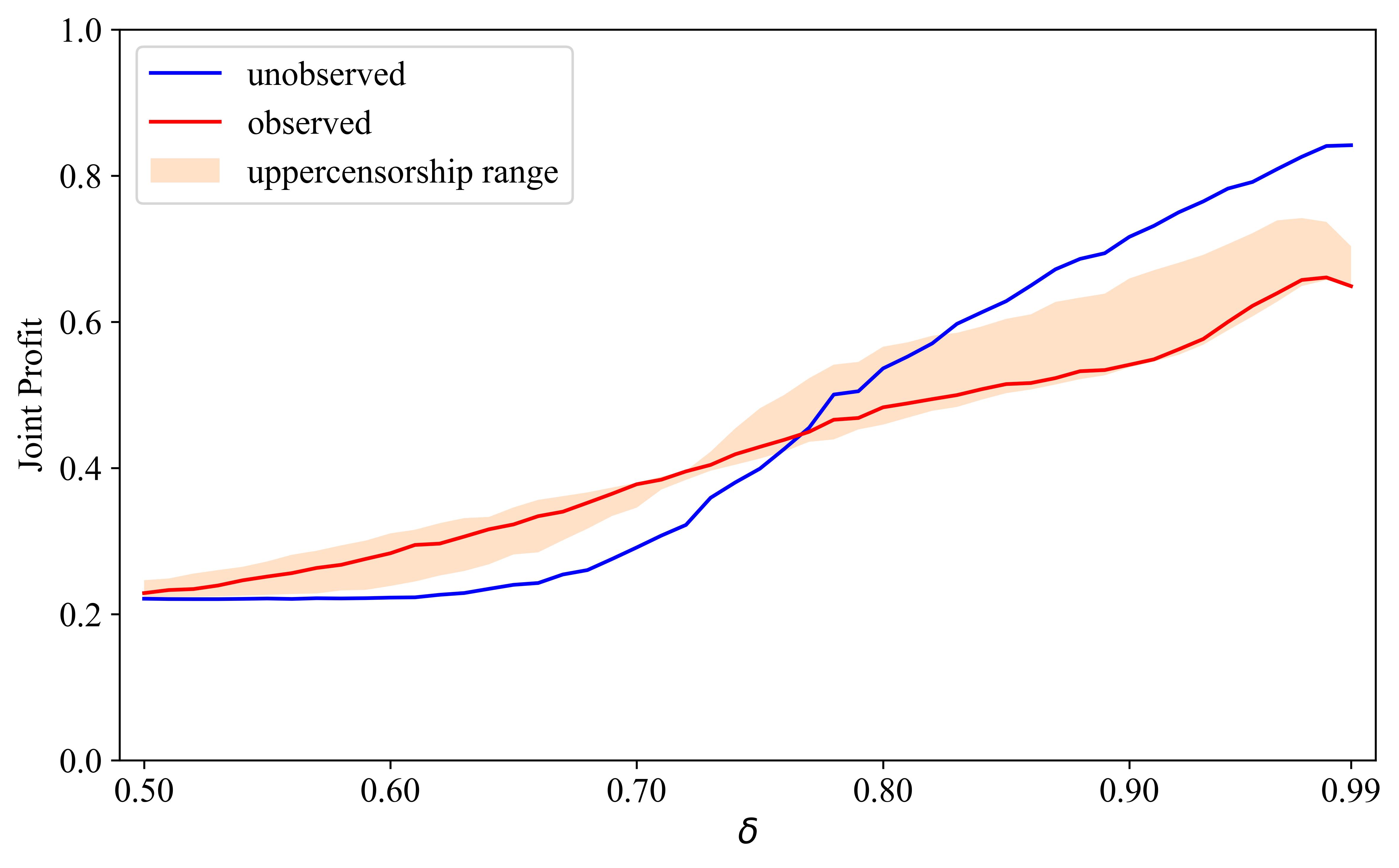}
    \caption{Joint Profit under Different Disclosure Rules}
    \label{fig:profit_delta}
\end{figure}

Figure~\ref{fig:profit_delta} displays the normalized joint profit under different disclosure rules across $\delta$. The blue line shows the profit path under no disclosure, while the red line corresponds to full disclosure. The pink region denotes the attainable profit area under upper censorship when varying the level of $\rho$. 
The general pattern is that profitability increases with $\delta$, showing that as agents become more patient and place greater weight on future continuation values, their expected profits rise.

Another important feature is that Q-learning agents indeed respond to how demand signals are released. That is, different information disclosure rules generate different effects on profits. In particular, Figure~\ref{fig:uppercensorship_delta} in the Appendix shows that under upper censorship, varying the truth-telling probability $\rho$ leads to different profit trajectories.
Moreover, the figure reveals a sharp cutoff in the optimal $\rho^*$: below this cutoff, releasing more information (a higher $\rho$) raises profits, whereas above it, pooling more information (a lower $\rho$) yields higher profits. This pattern contrasts sharply with the theoretical prediction that the optimal $\rho^*$ should increase with $\delta$.

\begin{flushleft}
\textbf{Result 1:} Q-learning agents earn higher expected profits as $\delta$ increases. In addition, they respond systematically to the structure of information disclosure.
\end{flushleft}

\begin{flushleft}
\textbf{Result 2:} Under upper censorship, varying the truth-telling probability $\rho$ leads to different profit paths. However, the optimal $\rho^*$ increases at low $\delta$ but decreases at high $\delta$, in sharp contrast with the theoretical prediction.
\end{flushleft}

We continue to examine upper censorship. The profit path under full disclosure always lies within the feasible profit region generated by upper censorship. In particular, relative to full disclosure, upper censorship expands the upper bound of attainable profits for all $\delta$. This result is fully consistent with the theoretical prediction that upper censorship dominates full disclosure, and it suggests that the dominance is even stronger in algorithmic pricing.

Comparison between no disclosure and full disclosure reveals a clear single-crossing pattern. Below the cutoff value of $\delta$, full disclosure generates higher profits than no disclosure, whereas above the cutoff, no disclosure yields higher profits than full disclosure. We refer to this pattern as \textit{profit reversal}. This reversal is the opposite of the theoretical prediction, which shows that no disclosure should yield higher profits at low $\delta$, while full disclosure should yield higher profits at high $\delta$.

\begin{flushleft}
\textbf{Result 3:} Upper censorship consistently delivers higher profits than full disclosure across the entire range of $\delta$.
\end{flushleft}

\begin{flushleft}
\textbf{Result 4:} The profit ordering between no disclosure and full disclosure exhibits a profit reversal: full disclosure outperforms no disclosure at low $\delta$, whereas no disclosure dominates at high $\delta$, exactly opposite to the theoretical prediction.
\end{flushleft}

\subsection{Discussion}
This profit reversal also aligns with the finding in \citet{ye2025algorithmic}, where the comparison between the observed demand shocks environment and the fixed-demand benchmark displays the same reversal pattern. Moreover, the simulation results under upper censorship show that when $\delta$ is low, disclosing more information (a higher $\rho$) increases profits, whereas pooling information (a lower $\rho$) reduces them. Taken together, these results point to a unifying insight about how Q-learning algorithms learn: more information can reduce profits when $\delta$ is low, while less information can enhance profits when $\delta$ is high. This highlights that the comparative statics produced by Q-learning can differ fundamentally from those implied by equilibrium reasoning.

\section{Robustness Check}\label{sec:robust}

\subsection{Selective Disclosure II: Signal Precision}\label{app:signalprecision}
To further evaluate upper censorship, we conduct a horse race with another canonical selective disclosure rule, namely mean-preserving noisy contraction on both demand states, referred to as \textit{signal precision}.
Under signal precision, the third party chooses the informativeness of the disclosed demand signal. Specifically, the third party sets a precision parameter $\rho \in (0.5,1)$, which determines the probability that the signal correctly reflects the true demand state. A higher value of $\rho$ produces a more informative signal, whereas a lower value introduces additional noise.
The signal structure and its posterior beliefs are summarized in Figure~\ref{fig:signalprecision}.

\begin{figure}[h!]
\centering
\begin{subfigure}{0.35\textwidth}
$$
\begin{array}{c|cc}
 & L & H \\ \hline
m=l & \rho & 1-\rho \\
m=h & 1-\rho & \rho
\end{array}
$$
\caption{Signal Structure}
\end{subfigure}
\hspace{0.03\textwidth}
\begin{subfigure}{0.35\textwidth}
$$
\begin{array}{c|cc}
 & L & H \\ \hline
m=l & \rho & 1-\rho \\
m=h & 1-\rho & \rho
\end{array}
$$
\caption{Posterior}
\end{subfigure}
\caption{Signal Precision in the Binary Case}
\label{fig:signalprecision}
\end{figure}

The expected demand conditional on receiving signals $l$ and $h$ are $\underline{\theta} = \rho \theta_L+(1-\rho) \theta_H$ and $\bar{\theta} = (1-\rho)\theta_L+\rho \theta_H$. 
The optimal precision parameter $\rho^*$ is the highest value for which the monopoly price associated with $\bar{\theta}$ remains incentive compatible.\footnote{This has the flavor of \citet{mt2019demand}. The difference is that their setting treats signal precision as exogenous, whereas here the precision is endogenously chosen through an information design problem to maximize profits.}

The simulation results show that upper censorship outperforms signal precision for all values of $\delta$, demonstrating the robustness of its superiority under algorithmic pricing.

\subsection{Logit Demand}\label{app:logitl} 
In the baseline, we find that Q-learning under linear demand exhibits a clear profit reversal between no disclosure and full disclosure. To further assess the robustness of this finding, we consider a more general Bertrand competition framework, which is the logit model that captures both horizontal and vertical differentiation.
There are two differentiated products supplied by the two agents, along with an outside good. In each period $t$, the demand for agent $i$ is given by
$$
D_{it}=\frac{e^{\frac{a_i-p_{it}}{\mu}}}{ 
e^{\frac{a_1-p_{1t}}{\mu}}+
e^{\frac{a_2-p_{2t}}{\mu}}+
e^{\frac{a_0}{\mu}}}
$$
where $a_i$ captures vertical differentiation and $\mu$ reflects the degree of horizontal differentiation. The parameter $a_0$ serves as an inverse index of aggregate market size. The period payoff of agent $i$ is $\pi_{it}=(p_{it}-c_i) q_{it}$, with $c_i$ being the marginal cost. To incorporate i.i.d. demand shocks, we allow the aggregate demand parameter $a_0$ to vary stochastically.
We follow the parameterization in \citet{calvano2020}, setting $c_i = 1$, $a_i - c_i = 1$, and $\mu = 1/4$. There are two demand shocks—a positive and a negative one—corresponding to $a_0^H = -0.15$ and $a_0^L = 0.15$, respectively. The price range spans from the Bertrand prices under $L$ to the monopoly prices under $H$, using the same grid size $m = 11$. We maintain a one-period memory ($K=1$).


The simulation results confirm that the profit reversal persists under logit demand: full disclosure yields higher profits at low $\delta$, while no disclosure performs better at high $\delta$. This rules out the demand environment as the source of the profit reversal.

\section{Conclusion}\label{sec:conclusion}
Firms increasingly acquire market information from third-party intermediaries. With the growing use of algorithmic pricing, the design of information disclosure rules has become crucial for understanding how pricing algorithms learn and compete under different disclosure environments. This paper examines how information disclosure rules shape the pricing behavior of Q-learning algorithms in markets with stochastic demand shocks. Using a unified simulation framework, we examine three disclosure rules—no disclosure, full disclosure, and upper censorship, which is the optimal rule for maximizing collusive profits—and evaluate their impact on learning outcomes.

Simulation results show that Q-learning algorithms respond systematically to the structure of information disclosure. Upper censorship consistently outperforms full disclosure for all discount factors, consistent with theoretical predictions from \citet{sugaya2025collusion}. We also document a robust profit reversal: full disclosure generates higher profits than no disclosure at low discount factors, whereas no disclosure yields higher profits at high discount factors. This pattern is exactly the opposite of the classical theoretical prediction. The reversal persists across demand environments, including both linear and logit demand, indicating that it is a structural feature of Q-learning rather than an artifact of the demand specification.

Our findings have important policy implications. Recent policy debates in the United States have proposed restricting or even prohibiting the use of nonpublic competitor data by third-party intermediaries, based on the assumption that limiting information sharing will help achieve antitrust objectives.
However, such policies may be misguided in the context of algorithmic pricing. In our simulations, reducing the informativeness of disclosure can increase profits when Q-learning agents are sufficiently patient, and no disclosure can even achieve the highest profits among all disclosure rules for high discount factors. These results indicate that restricting information sharing may backfire, potentially strengthening rather than weakening algorithmic collusion. Thus, we provide simulation evidence that complements the critique by \citet{harrington2025critique}, who argues that the source of competitive harm lies not in shared data.

To regulate algorithmic pricing, our findings suggest an alternative policy tool based on information design. Since Q-learning algorithms respond systematically to the structure of information disclosure, regulators can influence learning dynamics by shaping what information is revealed rather than simply releasing nothing. In some environments, learning outcomes align with theoretical predictions, while in others they diverge, which highlights that policy interventions must be evaluated not only through theory but also through simulation evidence, as emphasized by \citet{calvano2020protecting}.

\section*{Acknowledgement}
Computation reported in this work was carried out on the Unity Cluster of the College of Arts and Sciences at the Ohio State University and the HPC at the University of Arizona. The computational resources and support provided are gratefully acknowledged.

\newpage
\appendix
\section*{Appendix}

\renewcommand{\thesubsection}{\Alph{subsection}}
\renewcommand{\thefigure}{A.\arabic{figure}}
\setcounter{figure}{0}
\renewcommand{\thetable}{A.\arabic{table}}
\setcounter{table}{0}

\subsection{Tables}

\subsection{Figures}

\begin{figure}[H]
\centering
\begin{subfigure}{0.48\textwidth}
    \includegraphics[width=\textwidth]{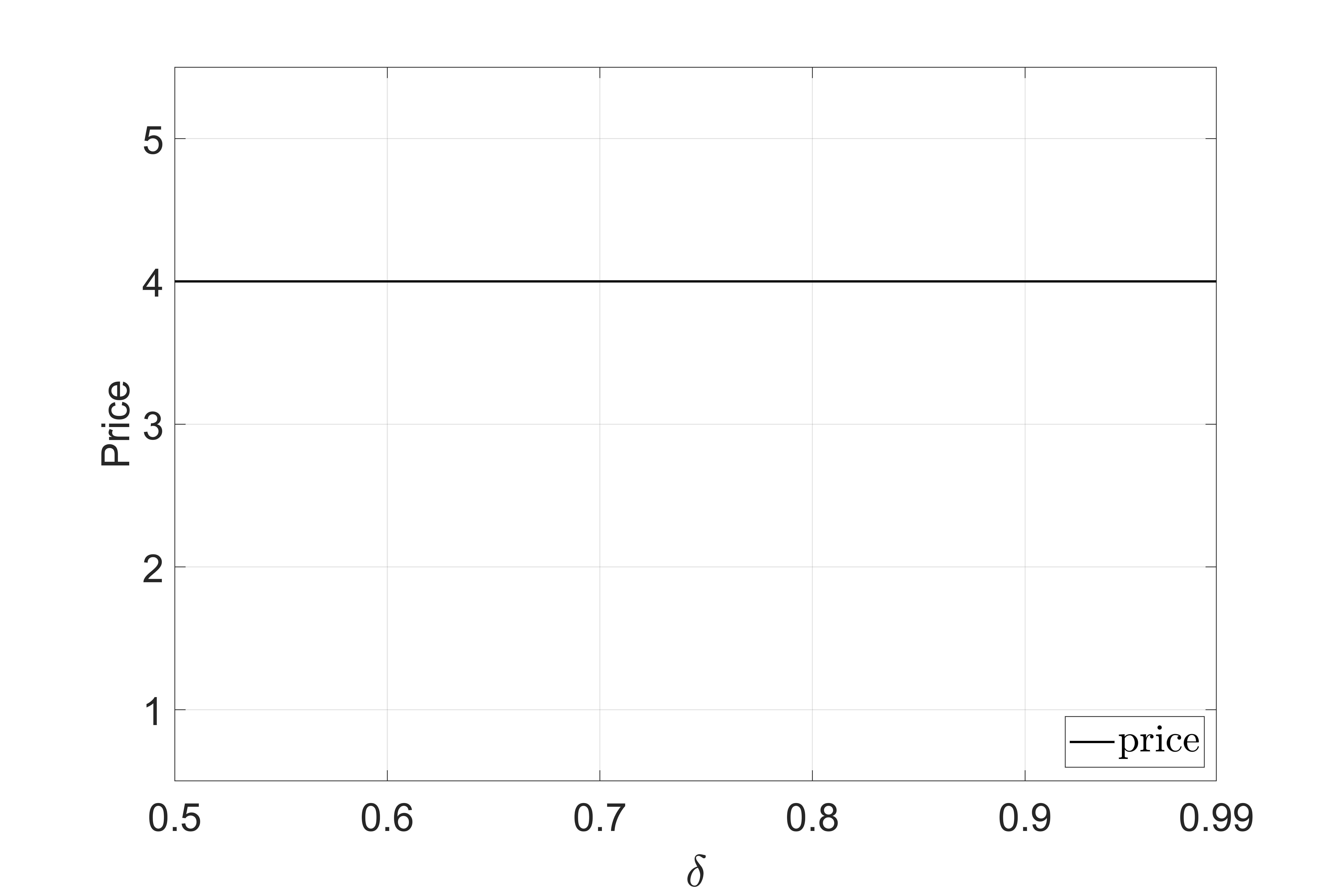}
    \centering
    \caption{No Disclosure}
    \label{fig:exampleG_1}
\end{subfigure}
\begin{subfigure}{0.48\textwidth}
    \includegraphics[width=\textwidth]{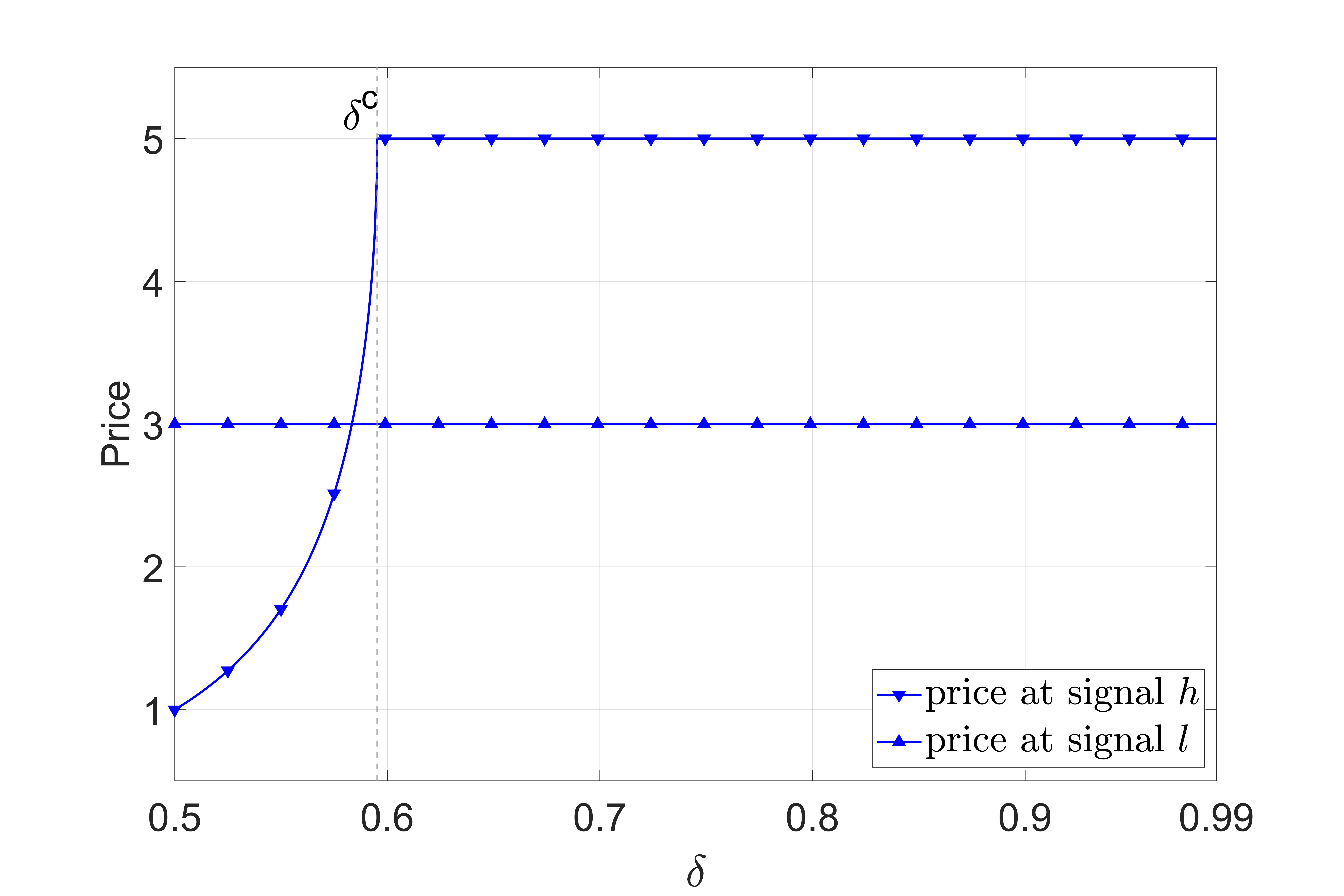}
    \centering
    \caption{Full Disclosure}
    \label{fig:exampleG_2}
\end{subfigure}
\hspace{2cm}
\begin{subfigure}{0.48\textwidth}
    \includegraphics[width=\textwidth]{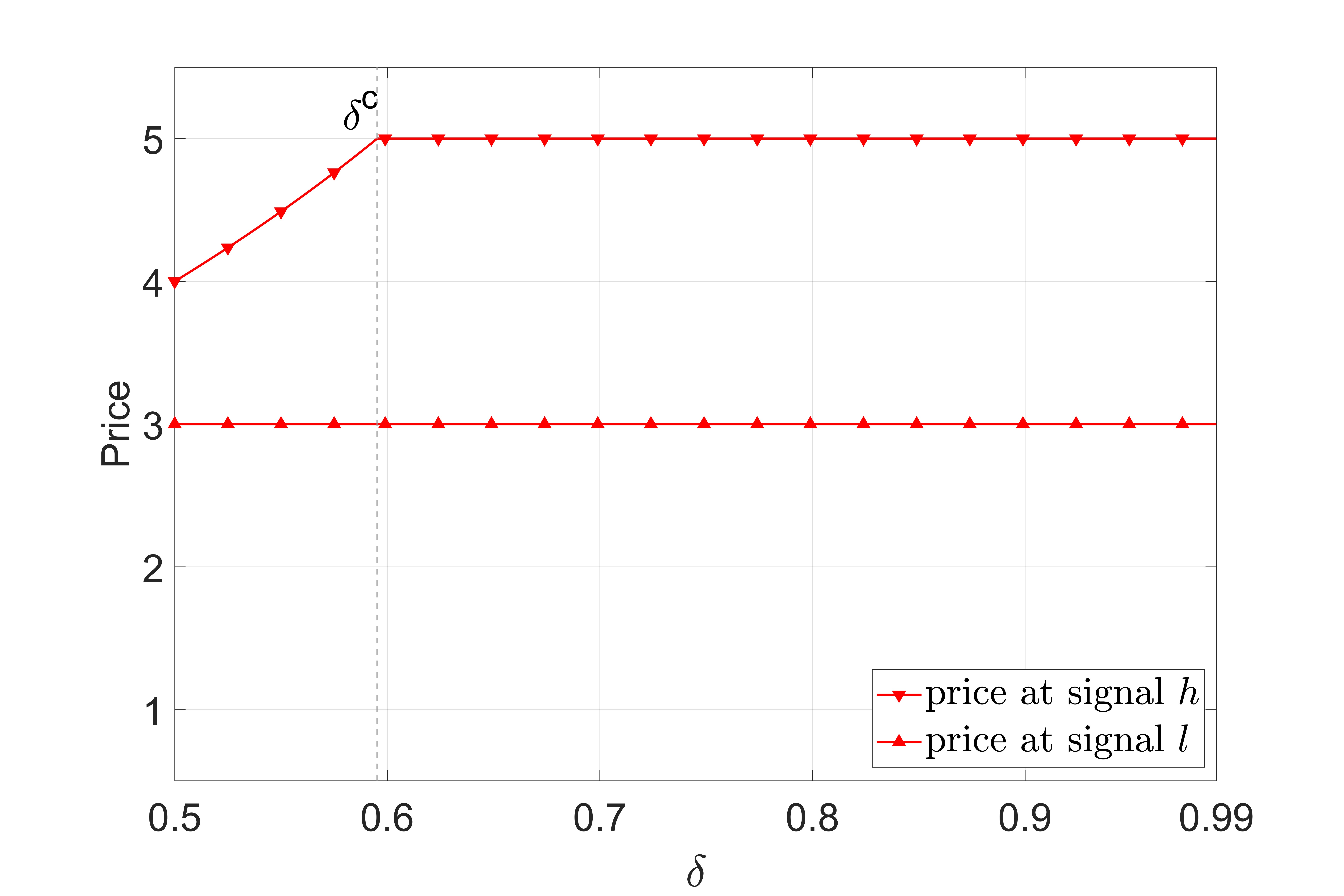}
    \centering
    \caption{Upper Censorship}
    \label{fig:exampleG_3}
\end{subfigure}
\begin{subfigure}{0.48\textwidth}
    \includegraphics[width=\textwidth]{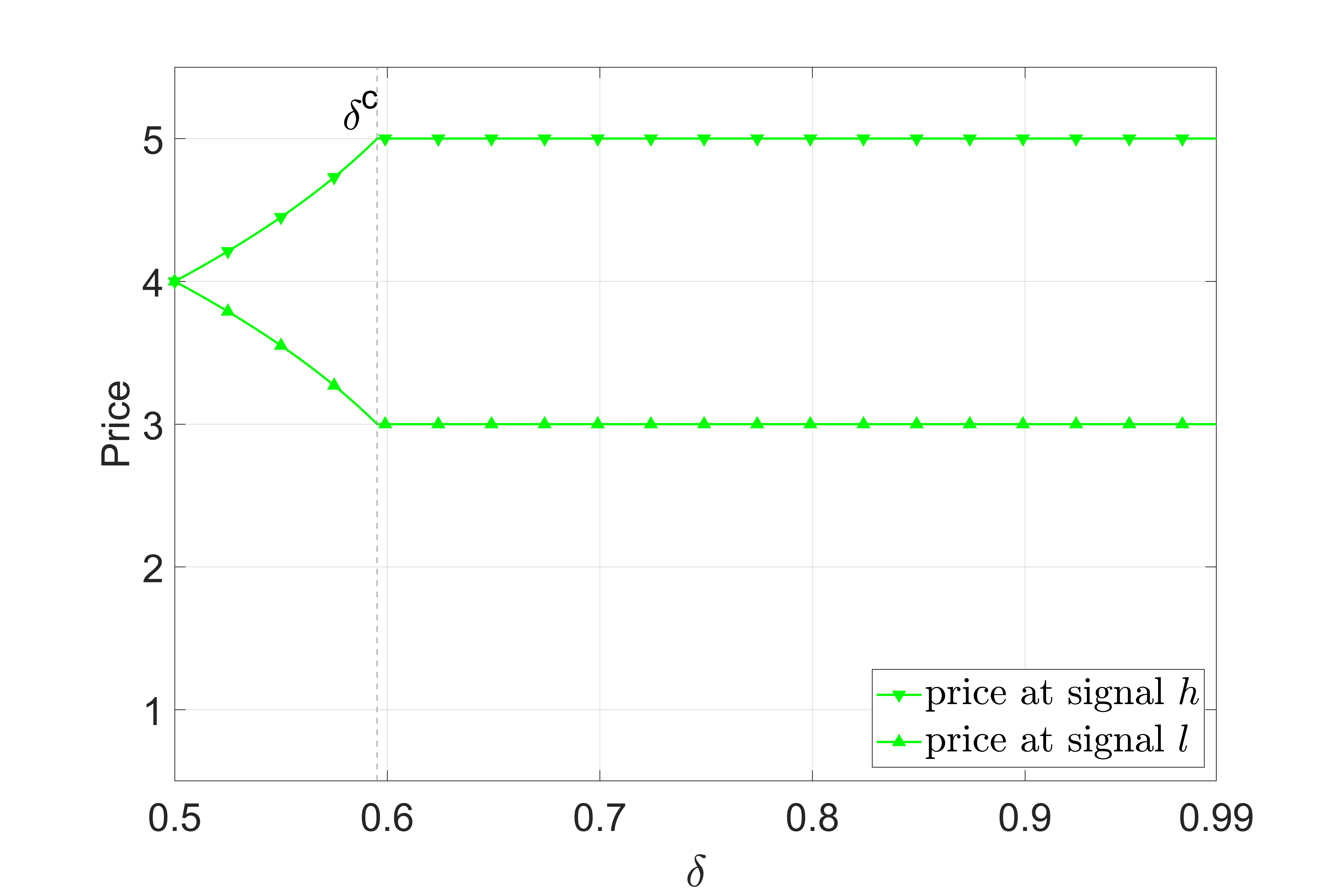}
    \centering
    \caption{Signal Precision}
    \label{fig:exampleG_4}
\end{subfigure}
\caption{Theoretical Optimal Prices under Different Disclosure Rules}
\label{fig:theory_prices}
\end{figure}

\begin{figure}[H]
    \centering
    \includegraphics[width=0.95\textwidth]{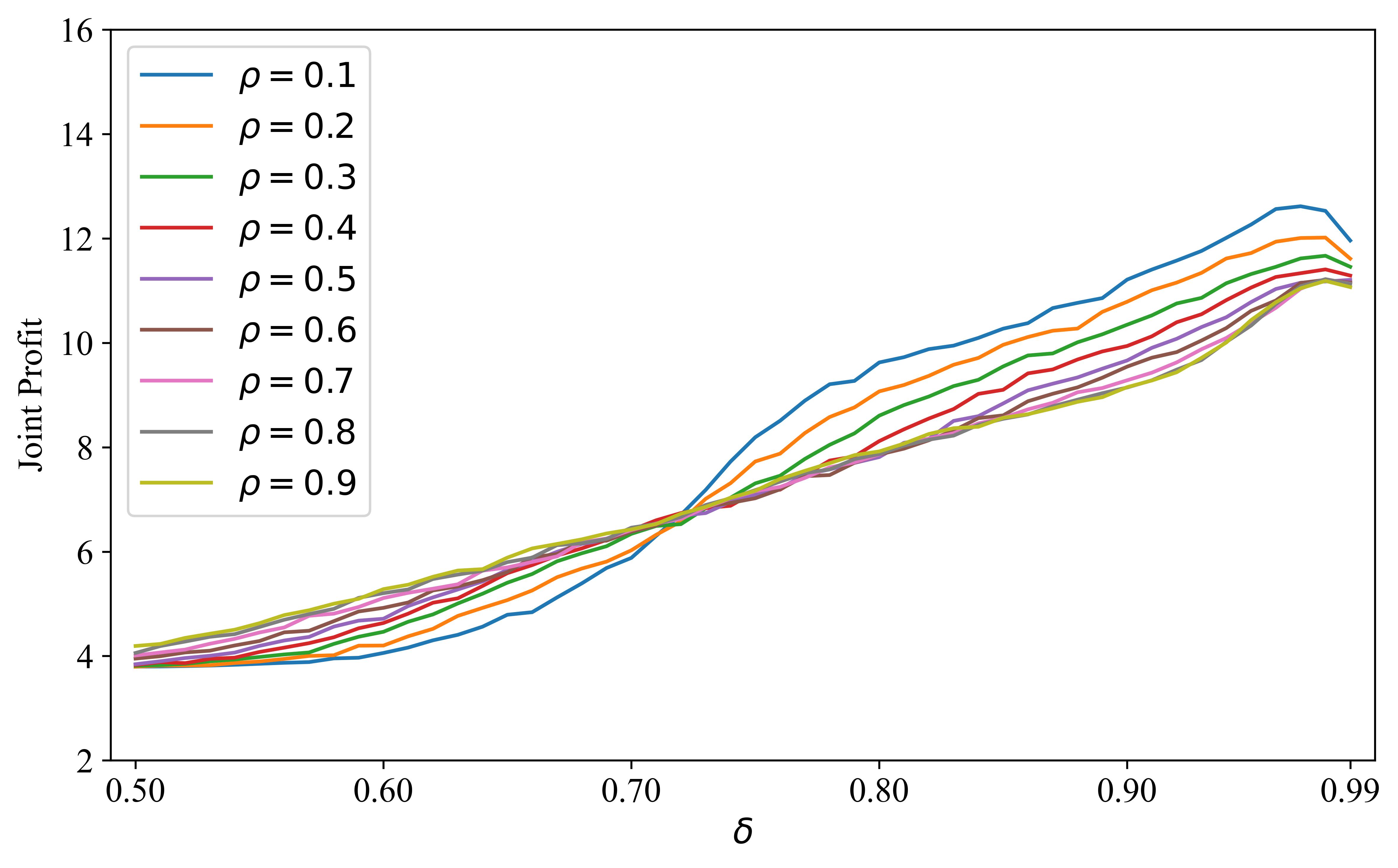}
    \caption{Joint Profit for Different Values of $\rho$ under Upper Censorship}
    \label{fig:uppercensorship_delta}
\end{figure}

\subsection{Technical Details}\label{app:techniques}
In this section, we mathematically show the optimal prices and optimal joint profits for each information disclosure rule.

\subsubsection{No Disclosure}\label{app:ND}
Under no disclosure, the firms maximize their expected profit based on the prior distribution. The expected joint profit is $$\frac{1}{2}\left(\theta_Hp_t - p_t^2\right) + \frac{1}{2}\left(\theta_Lp_t - p_t^2\right)$$ By taking derivative with respect to $p_t$, we have the first order condition $$\frac{1}{2}\left(\theta_H + \theta_L\right) = 2p^*$$ Therefore, the optimal price is $p^* = \frac{1}{4}(\theta_H + \theta_L)$, and the corresponding joint profit is $\Pi^* = \frac{1}{16}(\theta_H + \theta_L)^2$. The firms will not deviate if the optimal discounted profit is greater than the monopoly profit in one period $$\frac{\Pi^*}{2(1-\delta)} \geq \Pi^* \Longleftrightarrow \delta \geq \frac{1}{2}$$

\subsubsection{Full Disclosure}\label{app:FD}
When demand states are fully observable, the optimal price under $H$ is $p_H^* = \frac{1}{2}\theta_H$, and the optimal price under $L$ is $p_L^* = \frac{1}{2}\theta_L$. The expected optimal joint profit is $\Pi^* = \frac{1}{8}(\theta_H^2 + \theta_L^2)$. Since $\theta_L < \theta_H$, the fully collusive outcome can be sustained if the firms have no incentive to deviate under $H$ $$\frac{1}{4}\theta_H^2 \leq \frac{1}{8}\theta_H^2 + \frac{\delta}{2\left(1-\delta\right)}\Pi^* \Longleftrightarrow \delta \geq \delta^c = \frac{2 \theta_H^2}{3 \theta_H^2 + \theta_L^2}$$ When $\delta \in [0.5,\delta^c)$, the firms can play a constrained collusion outcome. They choose $p_L$ and $p_H$ to maximize the expected joint profit $$\frac{1}{2}\left(\theta_L - p_L\right)p_L + \frac{1}{2}\left(\theta_H - p_H\right)p_H$$ subject to the constraints that for $L$ and $H$, no firms have incentive to deviate $$\left(\theta_L - p_L\right)p_L \leq \frac{1}{2}\left(\theta_L - p_L\right)p_L + \frac{\delta}{4\left(1-\delta\right)}\left(\left(\theta_L - p_L\right)p_L + \left(\theta_H - p_H\right)p_H\right)$$ and $$\left(\theta_H - p_H\right)p_H \leq \frac{1}{2}\left(\theta_H - p_H\right)p_H + \frac{\delta}{4\left(1-\delta\right)}\left(\left(\theta_L - p_L\right)p_L + \left(\theta_H - p_H\right)p_H\right)$$ which can be written as $$\left(\theta_L - p_L\right)p_L \leq \frac{\delta}{2-3\delta}\left(\theta_H - p_H\right)p_H \text{ and } \left(\theta_H - p_H\right)p_H \leq \frac{\delta}{2-3\delta}\left(\theta_L - p_L\right)p_L$$ As above, the constraint under $H$ is binding. Note that for $\delta \in [0.5,\delta^c)$, $\frac{\delta}{2-3\delta} \geq 1$ always holds. Thus the constraint under $L$ is relaxed. Therefore, the optimal price under $L$ is $p_L^* = \frac{1}{2}\theta_L$ while the optimal price under $H$ solves $\left(\theta_H - p_H^*\right)p_H^* = \frac{\delta}{2-3\delta}\left(\theta_L - p_L^*\right)p_L^*$, which implies $p_H^* = \frac{1}{2}\left(\theta_H - \sqrt{\theta_H^2 - \frac{\delta}{2 - 3\delta}\theta_L^2}\right)$. The corresponding optimal joint profit is $\Pi^* = \frac{1 - \delta}{4(2 - 3\delta)}\theta_L^2$.

\subsubsection{Upper Censorship}\label{app:UC}
The third party commits to the information structure shown in Figure \ref{fig:uppercensorship} before observing the demand state. According to \citet{sugaya2025collusion}, this problem can be reduced to a static information design problem. By Section~\ref{app:FD}, full disclosure is optimal for all $\delta \geq \delta^c$, it remains to check the case where $\delta \in [0.5,\delta^c)$. Define the maximum profit compatible with incentives, given continuation value $V$, as $$\Pi^{max}\left(\delta, V\right) = \frac{\delta V}{1-\delta}$$ Given each demand state $\theta$, the third party is solving a static information design problem with payoff $$f_V(\theta) = \min\left\{\frac{1}{4}\theta^2, \Pi^{max}\left(\delta, V\right)\right\}$$ It is straightforward that $f_V(\theta)$ is convex in $\theta$ until it hits the cap $\Pi^{max}$, then constant. By Lemma 1 in \citet{sugaya2025collusion}, the optimal collusive profit $V^*$ is the greatest fixed point of the following equation $$V = \max_{G}\mathbb{E}_G\left[\min\left\{\frac{1}{4}\theta^2, \Pi^{max}\left(\delta, V\right)\right\}\right]$$ where $G$ is the mean preserving contraction of the prior, which is the firms' posterior upon observing the signal. By Figure~\ref{fig:uppercensorship}, the posterior mean is $x_l=\theta_L$ after signal $l$, and $x_h=\frac{(1-\rho)\theta_L + \theta_H}{2-\rho}$ after signal $h$. Given $V$, the third party wants $x_h$ as large as possible without forcing a price cut below monopoly price. So the optimal $x_h$ solves $$\frac{1}{4}x_h^2 = \frac{\delta V}{1-\delta}$$ Let $x^*=2\sqrt{\frac{\delta V}{1-\delta}}$, then the optimal choice is
\begin{itemize}
    \item if $x^* \geq \theta_H$, full disclosure;
    \item if $x^*<\theta_H$, choose $\rho$ so that $x_h=x^*$, i.e., $\rho^*=\frac{2x^*-\left(\theta_L+\theta_H\right)}{x^*-\theta_L}$.
\end{itemize}
The on-path pricing strategy is $p_l=\frac{1}{2}\theta_L$, and $p_h=\frac{1}{2}x^*$. It remains to solve the fixed point for explicit solution. Given the disclosure rule and on-path pricing strategy, the expected joint profit is $\frac{1}{4}\theta_L^2$ after signal $l$, and $\frac{\delta V}{1-\delta}$ after signal $h$. Note that the probability of sending signal $l$ is $\frac{\rho^*}{2}$, and the probability of sending signal $h$ is $1-\frac{\rho^*}{2}$. Therefore $$V^*=\frac{\rho^*}{2}\frac{\theta_L^2}{4} + \left(1-\frac{\rho^*}{2}\right)\frac{x^*}{4} = \frac{\theta_L^2 + \theta_L\theta_H + \left(\theta_H-\theta_L\right)x^*}{8}$$ Combine the two expressions for $V^*$, we have $$\frac{{x^*}^2}{4} = \frac{\delta}{1-\delta}\frac{\theta_L^2 + \theta_L\theta_H + \left(\theta_H-\theta_L\right)x^*}{8}$$ This is a quadratic in $x^*$: $$2(1-\delta){x^*}^2 - \delta\left(\theta_H-\theta_L\right)x^* - \delta\theta_L\left(\theta_L + \theta_H\right) = 0$$ By solving for the positive root, we have $$x^*=\frac{\delta\left(\theta_H-\theta_L\right) + \sqrt{\delta}\sqrt{8\theta_L^2 + 8\theta_L\theta_H + \delta\left(\theta_H^2 - 7\theta_L^2 - 10\theta_L\theta_H\right)}}{4\left(1-\delta\right)}$$ For simplicity, let $\Delta=\sqrt{\delta}\sqrt{8\theta_L^2 + 8\theta_L\theta_H + \delta\left(\theta_H^2 - 7\theta_L^2 - 10\theta_L\theta_H\right)}$, then $x^*=\frac{\delta\left(\theta_H-\theta_L\right) + \Delta}{4\left(1-\delta\right)}$. Therefore, the optimal price after signal $l$ is $p_L^*=\frac{1}{2}\theta_L$, while the optimal price after signal $h$ is $p_H^*=\frac{\delta\left(\theta_H-\theta_L\right) + \Delta}{8\left(1-\delta\right)}$, and the corresponding optimal joint profit is $\Pi^*=\frac{4\left(1-\delta\right)\left(\theta_L^2 + \theta_L\theta_H\right) + \delta\left(\theta_H - \theta_L\right)^2 + \Delta\left(\theta_H - \theta_L\right)}{32\left(1-\delta\right)}$.

\subsubsection{Signal Precision}\label{app:SP}
We also only need to focus on $\delta \in [0.5,\delta^c)$ since full disclosure is optimal for any $\delta\geq\delta^c$. Since the information structure is symmetric and the prior is $\frac{1}{2}$, the signals are sent equally likely: $\operatorname{Pr}(h)=\operatorname{Pr}(l)=\frac{1}{2}$. Then the posterior means are $$x_h=\theta_L + \rho\left(\theta_H-\theta_L\right), \text{ and } x_l=\theta_L+(1-\rho)\left(\theta_H-\theta_L\right)$$ As before, the candidate on-path prices, $p_h=\frac{x_h}{2}$, and $p_l=\frac{x_l}{2}$, are feasible only if incentive constraints are satisfied. Note that the maximum IC profit given continuation value $V$ is $\frac{\delta V}{1-\delta}$. Given $\rho>0.5$, the best symmetric equilibrium achieves the joint profit per signal $$\min\left\{\frac{x_s^2}{4},\frac{\delta V}{1-\delta}\right\},~s\in\{h,l\}$$ and $V$ must satisfy the fixed point $$V=\frac{1}{2}\min\left\{\frac{x_h^2}{4},\frac{\delta V}{1-\delta}\right\} + \frac{1}{2}\min\left\{\frac{x_l^2}{4},\frac{\delta V}{1-\delta}\right\}$$ Since $x_h \geq x_l$, the high signal is the one that can violate IC. If $\rho$ is too high, $x_h$ gets large and the high-signal monopoly profit $\frac{x_h^2}{4}$ can exceed the IC cap $\frac{\delta V}{1-\delta}$. Given the information structure, the third party's best choice is to pick the largest $\rho$ such that the high-signal monopoly profit is still IC. At the optimum, the high-signal IC constraints binds: $$\frac{\delta V}{1-\delta} = \frac{x_h^2}{4}$$ Under conditional monopoly pricing with no price cutting, $V=\frac{x_h^2+x_l^2}{8}$. Therefore, we have $$\frac{\delta}{1-\delta}\frac{x_h^2+x_l^2}{8} = \frac{x_h^2}{4} \Longleftrightarrow \delta x_l^2 = (2-3\delta)x_h^2$$ At the optimum, $x_l = \sqrt{\frac{2-3\delta}{\delta}}x_h$. Since signals are equally likely, we also have $x_h + x_l = \theta_H + \theta_L$. Hence $$x_h^* = \frac{\delta(\theta_L + \theta_H)}{\delta + \sqrt{\delta(2-3\delta)}}, \text{ and } x_l^* = \frac{\sqrt{\delta(2-3\delta)}(\theta_L + \theta_H)}{\delta + \sqrt{\delta(2-3\delta)}}$$ This pins down the optimal on-path prices $$p_h^*=\frac{\delta(\theta_L + \theta_H)}{2\left(\delta + \sqrt{\delta(2-3\delta)}\right)}, \text{ and } p_l^* = \frac{\sqrt{\delta(2-3\delta)}(\theta_L + \theta_H)}{2\left(\delta + \sqrt{\delta(2-3\delta)}\right)}$$ The corresponding optimal expected joint profit is $$\Pi^* = \frac{(1-\delta)\left(\theta_L + \theta_H\right)^2}{8\left(1-\delta+\sqrt{\delta(2-3\delta)}\right)}$$

For $\delta \leq \delta^c$, the optimal prices and joint profit under different information disclosure rules are summarized in Table~\ref{table:delta_price_profit}.

\begin{table}[h!]
\centering
\caption{Summary of Optimal Prices and Joint Profits under $\delta \in [0.5, \delta^c)$}

\scalebox{0.8}{ 
{\renewcommand{\arraystretch}{1.5}
\begin{tabular}{|c|c|c|c|}
\hline
\textbf{Information Structure} & $\mathbf{p_L^*}$ & $\mathbf{p_H^*}$ & $\mathbf{\Pi^*}$ \\ \hline

No Disclosure & $\frac{1}{4}(\theta_H + \theta_L)$ & $\frac{1}{4}(\theta_H + \theta_L)$ & $\frac{1}{16}(\theta_H + \theta_L)^2$ \\ \hline
Full Disclosure & $\frac{1}{2}\theta_L$ 
& 
$\frac{1}{2}\left(\theta_H - \sqrt{\theta_H^2 - \frac{\delta}{2 - 3\delta}\theta_L^2}\right)$ 
& 
$\frac{1 - \delta}{4(2 - 3\delta)}\theta_L^2$ \\ \hline

Upper Censorship & $\frac{1}{2}\theta_L$ 
& 
$\frac{\delta\left(\theta_H-\theta_L\right) + \Delta}{8\left(1-\delta\right)}$ 
& 
$\frac{4\left(1-\delta\right)\left(\theta_L^2 + \theta_L\theta_H\right) + \delta\left(\theta_H - \theta_L\right)^2 + \Delta\left(\theta_H - \theta_L\right)}{32\left(1-\delta\right)}$
\\ \hline

Signal Precision & $\frac{\delta(\theta_L + \theta_H)}{2\left(\delta + \sqrt{\delta(2-3\delta)}\right)}$ 
& 
$\frac{\sqrt{\delta(2-3\delta)}(\theta_L + \theta_H)}{2\left(\delta + \sqrt{\delta(2-3\delta)}\right)}$ 
& 
$\frac{(1-\delta)\left(\theta_L + \theta_H\right)^2}{8\left(1-\delta+\sqrt{\delta(2-3\delta)}\right)}$ 
\\ \hline

\end{tabular}}}
\label{table:delta_price_profit}

\end{table}


\clearpage
\setcitestyle{numbers} 
\bibliographystyle{chicago} 
\bibliography{refs} 

\end{document}